%% file: so5.tex
\input macros.tex

\input epsf.tex
%\draftmode

%\advanceclock{10}{32}

\nopageonenumber
\baselineskip = 18pt
\barsoff

% Local Definitions ---------------------------------------------------------

 2

\def\five{{$SO(5)$}}

\def\twothree{{$SO(3) \times SO(2)$}}

\def\veps{\varepsilon}

\def\bk{\item{}}

% Title page ----------------------------------------------------------------

%\rightline{DRAFT, July 24, 1997.}
\line{cond-mat/9707290 \hfil McGill-97/11, UdeM-GPP-TH-97-41}

\vskip .1in
\title
\centerline{A Microscopic Derivation of}
\centerline{the SO(5)-Symmetric}
\centerline{Landau-Ginzburg Potential}
\endtitle

\vskip 0.2in
\authors
\centerline{C.P.~Burgess${}^a$, J.M.~Cline${}^a$, R.~MacKenzie${}^b$ and
R.~Ray${}^b$}
\vskip .1in
\centerline{\it ${}^a$ Physics Department, McGill University}
\centerline{\it 3600 University St., Montr\'eal, Qu\'ebec, 
Canada, H3A 2T8.}
\vskip .05in
\centerline{\it ${}^b$ Laboratoire Ren\'e-J.-A.-L\'evesque,
Universit\'e de Montr\'eal}
\centerline{\it C.P. 6128, Succ. centre-ville, Montr\'eal, 
Qu\'ebec, Canada, H3C 3J7.}
\endauthors

\abstract
\vbox{\baselineskip 15pt
We construct a microscopic model of
electron interactions which gives rise to both
superconductivity and antiferromagnetism, and
which admits an approximate $SO(5)$ symmetry
that relates these two phases. The symmetry can be
exact, or it may exist only in the long-wavelength
limit, depending on the detailed form of the interactions.
 We compute the macroscopic Landau-Ginzburg
free energy for this model as a function
of temperature and doping, by explicitly integrating
out the fermions. We find that the resulting phase
diagram can resemble that observed for the cuprates,
with the antiferromagnetism realized as
a spin density wave, whose wavelength might be 
incommensurate with the lattice spacing away from
half filling.
\ignore
if the $SO(5)$ symmetry is weakly
violated in favor of the antiferromagnetic condensate, then 
the resulting phase diagram can have the experimentally 
desired properties: purely antiferromagnetic for small doping and 
superconducting for intermediate doping.
\endignore
}
\endabstract

% References ---------------------------------------------------------------

\ref\zhang{S.-C. Zhang, Science {\bf 275} (1997) 1089.}

\ref\greiter{M. Greiter, preprints cond-mat/9705049,
cond-mat/9705282.}

\ref\zhangreplies{S.C. Zhang, preprint cond-mat/9705191.}

\ref\anderson{G. Baskaran and P.W. Anderson, preprint cond-mat/9706076.}

\ref\similarmodel{S. Rabello, H. Kohno, E. Demler and
S.C. Zhang, preprint cond-mat/9707027.}

\ref\ourSOFivepapers{C.P. Burgess and C.A. L\"utken,
preprints cond-mat/9611070; cond-mat/9705216.}

\ref\joeandco{J. Polchinski,
{\it Effective Field Theory of the Fermi Surface}, in {\sl
Recent Developments in Particle Theory, Proceedings of the 1992 TASI},
eds. J. Harvey and J. Polchinski (World Scientific, Singapore, 1993);\bk
R. Shankar, {\it Renormalization Group Approach
to Interacting fermions}, \rmp{66}{94}{129}; \bk
T. Chen, J. Fr\"ohlich and M. Seifert,
{\it Renormalization Group Methods: Landau-Fermi Liquid
and BCS Superconductor}, preprint cond-mat/9508063.}

\ref\recentdata{P.-C. Dai, H.A. Mook and F. Dogan, cond-mat/9707112.}

\ref\lgcalcs{D.L. Feder and C. Kallin, \prb{}{97}{}, cond-mat/9609248;\bk
S. Stintzing and W. Zwerger, \prb{}{97}{}, cond-mat/9703129.}

\ref\Efns{I.S. Gradshteyn and I.M. Ryzhik, Table of Integrals, Series and
Products (Academic Press, New York, 1980)}

% Main text ----------------------------------------------------------------

\vfill\eject

\section{Introduction and Summary}

Zhang has recently proposed \zhang\ that the superconducting and
antiferromagnetic phases of the high-$T_c$ cuprates might be related to
one another by an approximate $SO(5)$ symmetry of the electronic
Hubbard Hamiltonian. This proposal incorporates in a fundamental way
the connection between antiferromagnetism (AF) and superconductivity 
(SC) which is observed in these systems.

Recently, however, the foundations of this picture have come under
attack, with criticism directed against the existence of an approximate
$SO(5)$ symmetry of the Hubbard Hamiltonian \greiter\ (for a reply, see
\zhangreplies), as well as against the very possibility, in principle,
of relating the antiferromagnetic and superconducting phases by the
rotation of a finite-dimensional order parameter \anderson.  This
motivates studying whether and under what conditions
an approximate $SO(5)$
symmetry can emerge from a microscopic picture of electron dynamics.

In this paper we investigate a model of electron dynamics which can
naturally incorporate an approximate $SO(5)$ symmetry.\foot\zhangsmodel{A
broader class of similar models has recently been 
proposed in ref.~\similarmodel.} 
Our model consists of degenerate 
electrons with two kinds of attractive interactions, 
described by a four-Fermi Hamiltonian of the form
$H = H_0 + H_{\rm int}$, where $H_{\rm int} = H_{\sss AF}
+ H_{\sss SC}$, and
\label\Hamform
\eqa
H_0 &= \sum_p \; (\veps_p-\mu) \; 
\psi_p^\dagger \psi_p + \sum_q \Bigl( a_\phi^0 \,
\phi_q^* \phi_q + a_n^0 \, \vec{n}_q^* 
\cdot \vec{n}_q \Bigr), \eol
H_{\sss AF} &= {1 \over 2V} \sum_{pq}  f(p,q) \; 
\Bigl( \psi_{p+Q+q}^\dagger \vec\sigma \, \psi_p \Bigr)
\cdot \vec{n}_{Q+q} + \hc, \eol
\label\Hamformthr
H_{\sss SC} &= {1 \over 2V} \sum_{pq}  g(p,q) \; 
\Bigl( \psi_{-p-q}^\sst \sigma_2 \, \psi_p \Bigr)
{\phi}_{q} + \hc, \eeol
\eeq
Here $\psi_p = \pmatrix{\psi_{p\uparrow} \cr 
\psi_{p\downarrow} \cr}$ is the electron field, while 
$\phi_q$ and $\vec{n}_q$ are Hubbard-Stratonovich auxiliary
scalar fields which, when integrated out, produce the two 
four-Fermi interactions. $\veps_p$ denotes the 
electron single-particle dispersion relation, $\mu$ is the 
chemical potential that measures deviations from half-filling,
and $\vec\sigma = \{ \sigma_i, i = 1,2,3 \}$ are the
Pauli matrices. $Q$ is a fixed nesting vector which will
be defined below.  $a_\phi^0$ and $a_n^0$ are positive constants which
could, if one wished, be absorbed into the definitions of $f$ and
$g$ by a redefinition of $\phi$ and $\vec n$; however it is 
convenient to keep them for the purposes of renormalization in
\S4. 

Our goal in this work is twofold: first, to find the conditions under
which this model exhibits an approximate \five\ symmetry, and to see how
the symmetry-breaking effects manifest themselves at temperatures much
less than the Fermi energy. We find the symmetry to be possible, 
and it can arise ``accidentally''at long wavelengths even if it is
not important for the microscopic dynamics. We also find that
symmetry-breaking effects are marginal, for weak couplings, in
the sense that they increase only logarithmically as one scales into the
infrared.  Second, we wish to compute the effective Landau-Ginzburg
(LG) potential, so as to determine how the phase diagram depends on the
microscopic electronic couplings.  The result is that the AF phase 
can have the phenomenologically desired shape in the 
temperature-doping plane, preferring low temperatures and
zero doping.  Moreover the SC phase, while also preferring
small doping, is not suppressed as quickly by increased doping as is
the AF phase.  The result is that SC becomes energetically favorable to
AF at some lower critical doping, and persists until some upper
critical doping. 
%A mixed phase, consisting of both superconductivity
%and antiferromagnetism, is never favoured. 
These results are
consistent with but more predictive than
the general effective-field-theory
description of \zhang\ and \ourSOFivepapers.
For example they predict the absence of a mixed AF/SC phase away
from zero doping when the fundamental Hamiltonian is \five-invariant. 
They also agree with the observed phase diagram of 
the high-$T_c$ systems, although we do not attempt
here a detailed discussion of the applicability 
of these models to the cuprates.

Our analysis of the model defined by eqs.~\Hamform--\Hamformthr\ is
motivated by the renormalization-group (RG) approach to understanding
superconductivity and antiferromagnetism as BCS and spin-density-wave
instabilities within Fermi liquid theory \joeandco.  In the RG
language, a Fermi liquid is understood as a regime of scales for which
the dominant quasiparticle degrees of freedom are weakly-coupled,
degenerate fermions carrying the same quantum numbers as the underlying
electrons. Under such circumstances almost all of the quasiparticle
self-interactions are irrelevant in the RG sense: they become less and
less important as one integrates out high energy modes to obtain an
effective Hamiltonian valid near the Fermi surface.

The only exceptions to the rule that interactions are irrelevant in the
infrared are certain four-Fermi terms, which can be marginally relevant
for special kinds of electron kinematics. An important example is any
pairwise attraction between electrons (or holes) having opposite
momenta. This kind of interaction %is marginally relevant, \ie\ it
grows logarithmically in the infrared, eventually becoming strong
enough to trigger the BCS instability at sufficiently
low energies. A second exception exists when the quasiparticle Fermi
surface (FS) is nested.\foot\nestingdef{The FS is nested if its opposite
edges are related to one another by a fixed tranlation, $\ss Q$,
in momentum space.} In this case the attraction between electrons
whose momenta sum to the nesting vector, $Q$, is also marginally
relevant, potentially triggering an instability towards the formation
of a condensate which is modulated in space with wavevector
$Q$. The result is a charge- or spin-density wave depending on
the electric charge and spin of the attractive channel. 

We have both types of instability in mind when using
the Hamiltonian in eqs.~\Hamform--\Hamformthr. It is assumed that the
quasiparticle energies $\veps_p$ are time-reversal invariant,
$\veps_{-p} = \veps_p$, and have a FS with a nesting vector $Q$, 
defined by the property $\veps_{p+Q} = -\veps_p$ for 
$p$ near the FS. (Our convention is to define the FS at half filling
to be the zero of energy.) 
An example of a dispersion relation with these properties
in two dimensions is that of the extended Hubbard model,
\label\dispex
\eq
\veps_p = -2t \Bigl[ \cos(p_x a)
+ \cos(p_y a) \Bigr] , 
\eeq
where $a$ is the lattice spacing and the nesting vector is $Q = \left(
\pm\, {\pi \over a}, \pm \, {\pi \over a} \right)$.  However the general 
analysis that follows in \S4 does not depend on using this 
explicit form.  Notice that 
because antiferromagnetism arises here as a spin-density wave, its
periodicity need not be precisely commensurate with the lattice
spacing, particularly away from half filling. %Although our model
%was not motivated by these experiments, 
Such incommensurate
antiferromagnetism is particularly interesting in view of 
the splitting of the AF peaks which is seen in
recent neutron-scattering data \recentdata.

Since eqs.~\Hamform--\Hamformthr\ involve an attractive interaction in
the spin-triplet channel, the nesting produces a spin-density-wave
instability. The Hubbard-Stratonovich formulation of the Hamiltonian,
eqs.~\Hamform--\Hamformthr, is particularly convenient for following
these instabilities of the system since they may be studied by
examining the minima of the potential for the modes $\phi_0$ and
$\vec{n}_Q$, which is induced when the electrons are integrated out.
Depending on which of these modes condenses most strongly, the
resulting ordered phases are superconductors or antiferromagnets,
possibly both.

Our presentation is organized as follows. In the next section, \S2, the
conditions for the Hamiltonian, eqs.~\Hamform--\Hamformthr, to exhibit
\five\ symmetry are derived. We find that the symmetry is present in
the long-wavelength limit, $q \to 0$, if the coupling functions satisfy
the relation $f^2 = g^2$. In \S3 we integrate out the electrons to
obtain the long-wavelength effective potential for the order
parameters, $\phi_0$ and $\vec{n}_Q$. The expressions obtained in \S3
are evaluated in \S4, in the limit of degenerate electrons, \ie\ with
all couplings and dispersion relations linearized around the FS. We
present explicit expressions in this limit for the coefficients of
the Landau-Ginzburg theory, as functions of the fundamental fermion
couplings, temperature and chemical potential. In some parts of the phase
diagram the usual expansion of the free energy to quartic order in the
fields is unbounded from below, forcing us to use the
full potential. From these results it is straightforward to determine
the main features of the phase diagram.  We find the dependence on $T$
and $\mu$ can have the desired form if the AF interaction in eq.~(2)
is chosen to be somewhat ($\sim 10-20\%$) stronger than the SC
interaction of eq.~(3).  In \S5 we examine the validity of the
degenerate-electron approximation by recomputing the quadratic and
quartic terms of the effective potential without linearizing about the
FS.  We do so using specific choices of $\veps$, $f$ and $g$ that have
been previously suggested [1,5,9] in connection 
with $d$-wave superconductivity, to show that the approximations used in 
\S4 are indeed reliable.  \S6 concludes with a brief recapitulation of 
our results.

\vfill\eject
\section{The $SO(5)$ Symmetry}

In this section we identify the conditions under which
the Hamiltonian, eqs.~\Hamform--\Hamformthr, admits an \five\ symmetry
along the lines Zhang has proposed.  

\subsection{The Algebra}

\five\ symmetry 
contains as a subgroup the \twothree\ group of spin 
rotations and electromagnetic gauge transformations, with
respect to which $H$ is obviously invariant. The electron
part of the corresponding
conserved charges (electric charge and spin) are given by:
\label\QandS
\eq
{\cal Q} = \sum_p \psi_p^\dagger \, \psi_p, \qquad \hbox{and}
\qquad 
\vec{S} = \hf \sum_p \psi_p^\dagger \vec\sigma \psi_p .
\eeq

The nontrivial requirement for \five\ invariance is the 
existence of the electrically-charged, spin-triplet 
off-diagonal operators, $\vec\Pi$, whose real and 
imaginary parts fill out the generators of \five\ in 
addition to ${\cal Q}$ and $\vec S$. Motivated by Zhang's \zhang\
choice, we take its action on electrons to be given by
\label\pidef
\eq
\vec\Pi \equiv \sum_p h_p \; (\psi^\sst_{-p+Q} \sigma_2 
\, \vec \sigma \, \psi_p ),
\eeq
where $h_p$ is a function to be determined. As is easily
verified, this has the desired commutation relations with
${\cal Q}$ and $\vec S$, and satisfies $\Bigl[ \Pi_a , \Pi_b
\Bigr] = 0$. In addition, if the function $h_p$ shares 
the same properties as does $\veps_p$: \ie\
$h_{p + Q} = - h_p$, and $h_{-p} = h_p$, then: 
\label\pipirelation
\eq
\Bigl[ \Pi_a , \Pi_b^* \Bigr] = 4 \sum_p \;
\left| h_p \right|^2 \left[ ( \psi^\dagger_p \, 
\psi_p ) \; \delta_{ab} -i \epsilon_{abc} \; 
( \psi^\dagger_p \, \sigma_c \, \psi_p) \right].
\eeq
Eq.~\pipirelation\ shows that $\{{\cal Q}, \vec S, \vec \Pi,
\vec \Pi^* \}$ satisfy the \five\ commutation relations 
provided we choose $|h_p| = c$, where $c$ is an 
appropriately chosen constant. This is only consistent
with the property $h_{p+Q} = - h_p$ if $h_p = \pm c$ 
for $p$ inside the FS, and $h_p = \mp c$ outside. 

\vfill\eject
\subsection{Conditions for Symmetry}

We next determine the conditions under which the Hamiltonian,
eqs.~\Hamform--\Hamformthr, commutes with the $\vec\Pi$.  
A nonvanishing chemical
potential explicitly breaks \five\ down to $SO(3)\times SO(2)$,
so we set $\mu=0$ here.  Furthermore, we ignore for now the scalar 
part of $H_0$, which is invariant provided $\phi$ and $\vec n$ 
transform together in the fundamental representation of \five.  
The electron part of the free Hamiltonian then satisfies:
\label\freecommutator
\eq
\Bigl[ \Pi_a , H_0^{el} \Bigr] = -2 \sum_p \veps_p \, h_p \;
( \psi^\sst_{Q-p} \, \sigma_2 \sigma_a \, \psi_p ) .
\eeq
This vanishes by virtue of the Fermi statistics of $\psi$,
provided that $\veps_p$ shares the previously advertised
symmetry properties of $h_p$: \ie\ if $\veps_{p+Q} = - \veps_p$ and 
$\veps_{-p} = \veps_p$. 

For notational convenience, we
define the operators appearing in $H_{\rm int}$
by $V_b(q)$ and $W(q)$, where $V_b(q) \equiv
\sum_p f(p,q) \; (\psi^\dagger_{p+q+Q}
\, \sigma_b \, \psi_p )$, and $W(q) \equiv 
\sum_p g(p,q) \, ( \psi^\dagger_{-p+q} \, \sigma_2 
\, \psi_p^*)$. Then
\label\comrelations
\eq
\eqalign{
\Bigl[ \Pi_a , W^*(q) \Bigr] &= 2 \sum_p \Bigl[
g^*(p+q-Q,q) + g^*(Q-p,q) \Bigr] \, h_p
\; (\psi^\dagger_{p-Q+q} \, \sigma_a \, \psi_p) , \cr
\Bigl[ \Pi_a , V_b(q) \Bigr] &= 2 \sum_p f(p,q)
\, h_{p+q} \; (\psi^\sst_{-p-q} \, \sigma_2 \, \sigma_a
\sigma_b \, \psi_p )\cr
&=  \sum_p \Bigl[f(p,q) \, h_{p+q} + f(-p-q,q) \, h_{-p}
\Bigr] \; (\psi^\sst_{-p-q} \, \sigma_2 \, \psi_p )\cr
& \qquad\qquad +
i \epsilon_{abc} \sum_p \Bigl[f(p,q) \, h_{p+q} - f(-p-q,q) \, h_{-p}
\Bigr] \; (\psi^\sst_{-p-q} \, \sigma_2 \, 
\sigma_c \, \psi_p ) . \cr}
\eeq
The operators $V_a$ and $W$ rotate into one another as
a five-dimensional vector of \five\ provided the coupling
functions satisfy the conditions 
$f(p,q) \, h_{p+q} \equiv f(-p-q,q) \, h_{-p} \propto g^*(p,q)$
and $[g^*(p+q-Q,q) + g^*(Q-p,q)] h_p \propto f(p,q)$.   

Of particular interest for many purposes, particularly the
phase diagram, is the long-wavelength limit, $q \to 0$. In
this limit we have $\Bigl[ \Pi_a , V_b(0) \Bigr] = 
-4 \delta_{ab} \, W^*(0)$ and $\Bigl[ \Pi_a , W(0) \Bigr] 
= -4 \; V_a(0)$, provided the functions 
$f_p \equiv \lim_{q \to 0} f(p,q)$ and 
$g_p \equiv \lim_{q \to 0} g(p,q)$ satisfy the properties
$f_p=f_{-p}$, $g_p =g_{-p} = - g_{p+Q}$ and
$f_p h_p = g_p$ and $g_p h_p = f_p$. Given our earlier choice that
$h_p = \pm c$, with opposite signs inside and outside the FS, we see
that $W(0)$ and $V_b(0)$ fill out the fundamental representation of
\five\ provided that the coupling functions satisfy $f_p = {\rm sign}(\veps_p)
 g_p$.

Finally, so long as $W$ and $V_b$ transform as a {\bf 5} of \five, the
entire Hamiltonian, eqs.~\Hamform--\Hamformthr, is \five\ invariant,
provided that \five\ is taken also to rotate the five scalars,
Re($\phi_0)/\sqrt{2}$, Im($\phi_0)/\sqrt{2}$ and $\vec{n}_Q$, into one
another as a {\bf 5}. Notice that although the demands made on $f$ and
$g$ for exact \five\ symmetry are quite strong, the symmetry can emerge
for small $q$ under much weaker conditions, and this is sufficient for
the invariance of the phase diagram.

\section{Integrating out the Electrons}

We now turn to the derivation of the effective theory which governs the
long-wavelength scalar modes, $\phi_0$ and $\vec{n}_Q$. The resulting
effective potential will be subsequently used to identify the phase
diagram of the system, and to see to what extent it exhibits
\five\ invariance.  Since the effective potential must be symmetric
under $U(1)$ transformations of $\phi_0$ and under $SO(3)$ rotations of
$\vec{n}_Q$, for notational simplicity we shall take $\phi_0\equiv\phi$
real and $(n_Q)_i=n \, \delta_{i,3}$.

\subsection{Preliminary Issues}

The thermodynamic variables we wish to follow in the LG potential are
the values of the fields $\phi$ and $n$ themselves, as well as the
temperature, $T$, and the electron doping away from half-filling, $x$.
We therefore ignore all scalar modes apart from $\phi$ and $n$, and
compute the effective potential for these variables.  In order to
follow the doping we also introduce a chemical potential, $\mu$, which
measures the change in density away from the free-particle FS (which is
defined here by $\veps_p = 0$).

We now integrate out the electrons, using the 
Hamiltonian, eqs.~\Hamform--\Hamformthr, supplemented by the
chemical potential. To incorporate the Fermi sea
we exchange the electron field, 
$\psi_p$ for $p$ inside the FS, for a hole field, 
$\chi_p = \sigma_2 \, \psi_{-p}^*$, in the usual way,
permitting the use of vacuum propagators for
the integration over the fermion fields. 

For the purposes of performing the functional integral
over $\psi_p$ and $\chi_p$,  it is
convenient to group $\psi_p$, $\psi_p^*$ and their
counterparts having momentum $-p$ and $p-Q$ into 
the following 16-component vector:
\label\sxtnvec
\eq
\Psi_p \equiv \pmatrix{ \psi_p \cr \psi_p^* \cr
\psi_{p-Q} \cr \psi_{p-Q}^* \cr  \psi_{-p} \cr 
\psi_{-p}^* \cr \psi_{Q-p} \cr \psi_{Q-p}^* \cr } =
\pmatrix{ \psi_p \cr \psi_p^* \cr
\sigma_2 \, \chi^*_{Q-p} \cr \sigma_2 \, \chi_{Q-p} \cr 
\psi_{-p} \cr \psi_{-p}^* \cr \sigma_2 \, \chi^*_{p-Q} \cr 
\sigma_2 \, \chi_{p-Q} \cr} .
\eeq
When $\Psi_p$ is used as the field, all sums over $p$
must be restricted to the original Brillouin zone, modulo
the transformations $p \to -p$ and $p \to p+Q$.  

With this choice the quadratic part of the Hamiltonian
may be written $\hf \; \Psi_p^\dagger \, \Delta \,
\Psi_p$, with:
\label\deltadef
\eq
\Delta \equiv \pmatrix{ A & B \cr B & A \cr},
\eeq
and $A$ and $B$ denoting the following $8\times 8$
matrices:
\label\Adef
\eq
A = \pmatrix{ 
\veps_p - \mu & 0& f_p n \, \sigma_3 & 0\cr
0& - \veps_p + \mu & 0& f_p n \, \sigma_3 \cr
f_p n \, \sigma_3 &0 & - \veps_p - \mu &0 \cr
0 & f_p n \, \sigma_3 & 0& \veps_p + \mu \cr} ,
\eeq
and
\label\Bdef
\eq
B = \pmatrix{ 
 0& g_p \phi \, \sigma_2 & 0& 0\cr
  g_p \phi \,\sigma_2 & 0& 0& 0\cr
0& 0 & 0& g_p \phi \, \sigma_2\cr
0 & 0& g_p \phi \,\sigma_2 &0 \cr} .
\eeq
In these last expressions, each entry of the matrices
$A$ and $B$ are $2\times 2$ matrices in spin space.

\subsection{The Functional Determinant}

Standard methods may now be used to compute the functional
integral over the fermion fields. Using the matrix
form for the Hamiltonian, eqs.~\deltadef--\Bdef, 
the result for the one-loop, finite-temperature contribution 
to the free energy density, $F_1$, in $d$ spatial dimensions is:
\label\FEresult
\eq
\eqalign{ -\beta F_1(\phi,n) & 
= \; \hf \, \ln \det \Bigl[- \partial_t^2 + \Delta^2\Bigr] \cr
&=  \hf \int {d^dp \over
(2 \pi)^d} \sum_{j=-\infty}^\infty \Bigl[ \ln \Bigl(
\omega_j^2 + \lambda_+^2 \Bigr) + \ln \Bigl(
\omega_j^2 + \lambda_-^2 \Bigr) \Bigr], \cr
&=   \int {d^dp \over
(2 \pi)^d} \; \ln \left[ {\cosh \left(  \beta \,
\lambda_+ /2\right) \cosh \left(  \beta \,
\lambda_- /2\right) 
} \right], \cr}
\eeq
where, as usual, $\beta = 1/kT$ and the sum is over 
the Matsubara frequencies, $\omega_j = (2j + 1) \pi \beta$.
We have switched to continuum momenta, and restored the 
integration region to the full Brillouin zone (so that it is
no longer modded out by $p \to -p$ and $p \to p + Q$).
The eigenvalues of the Hamiltonian $\Delta$ are given by
\label\evalresult
\eq
\lambda_\pm = \Bigl[ \Bigl( \sqrt{ \veps_p^2 + (f_p n)^2}
\pm \mu \Bigr)^2 + (g_p \phi)^2 \Bigr]^{\hf} .
\eeq

Even before undertaking a detailed evaluation of the free energy
\FEresult, we can prove an interesting result: in the \five\ limit
where $f_p^2/a^0_n = g_p^2/a^0_\phi$, 
and for nonzero doping, $\mu \ne 0$, the ground state must have $\phi=0$
or $n=0$ (possibly both) and hence is never a mixed phase of AF and SC.
This follows from the form of the energy eigenvalues \evalresult\ in
the conditions for minimizing the free energy.  Adding $F_1$ to the
tree-level free energy $F_0 = a_\phi^0\phi^2 + a_n^0n^2$, 
these conditions are
\eq
	\frac12{\partial F\over \partial X} = a_X^0 X - \hf
	\int {d^dp \over (2 \pi)^d}\sum_\pm 
	\tanh \left(\beta \, \lambda_\pm/2 \right) 
	{\partial \lambda_\pm\over \partial X}
\eeq
where $X$ stands for either of the fields $\phi$ or $n$. The 
derivatives of the eigenvalues have the form
\eq
	{\partial \lambda_\pm\over \partial \phi} = 
	{g_p^2\phi\over \lambda_\pm};\qquad 
	{\partial \lambda_\pm\over \partial n} = 
	{f_p^2 n\over \lambda_\pm}\left(1 \pm 
	{\mu\over\sqrt{\veps_p^2+(f_p n)^2}}\right)
\eeq
When $f_p^2=g_p^2$
and $a_\phi^0=a_n^0=a^0$, the minimization conditions can 
be written in the form
\eq
	\hf \, {\partial F\over \partial \phi} = \phi(a^0 - G), \qquad
\hf \, 	{\partial F\over \partial n} = n(a^0 - G + \mu^2 H),
\eeq
where it is easily shown that $G$ and $\mu H$ are positive definite functions.
Therefore in this limit there are never simultaneous solutions where both
$\phi$ and $n$ are nonvanishing, except when $\mu$ vanishes.

When evaluating these expressions in the next section,
% we proceed in two ways. First, 
we focus on the degenerate limit, for which $kT$ is much
less than the Fermi energy at zero doping. For example,
using the explicit dispersion
relation given by eq.~\dispex, this corresponds to
neglecting powers of $kT/t$. In this limit only states near the
Fermi surface are important in the integrals, and 
it is useful to decompose the momentum $p$ into a component $k$
on the FS and a component $\ell$ perpendicular to the
FS. We may then linearize all quantities in $\ell$, 
restrict the coupling functions $g$ and $f$
to momenta lying in the FS, and obtain
for the dispersion relation $\veps \approx v_\ssf \ell$, 
with the Fermi velocity, $v_\ssf(k)$, generally being a 
function of momentum, $k$. 

Eqs.~\FEresult\ and \evalresult\ are this section's 
main results. We now extract their implications for
the LG potential,\foot\lgrefs{See refs. \lgcalcs\
for related calculations of the LG free energy for
high-$\ss T_c$ systems.} first by examining them in detail in the
vicinity of zero fields, $\phi = n = 0$,
and then by computing the potential for arbitrarily values
of the fields, in certain directions in the $\phi$-$n$
plane.

\section{The Landau-Ginzburg Theory}

Eqs.~\FEresult\ and \evalresult\ define the effective potential for the
SC order parameter, $\phi$, and the AF order parameter, $n$.
Since this potential is analytic \foot\ztsingular{The same is not true
of the zero-temperature limit of the potential, which has the
nonanalytic behavior $\ss \phi^2 \ln \phi^2$ in the vicinity
of $\ss \phi = 0$.} at $\phi = n = 0$, an expansion to quartic
order in the fields gives a good indication of when a condensate will
form for either $\phi$ or $n$. However we will find that
the quartic terms do not always stabilize the potential, and 
it is necessary to know how the full potential looks away from 
$\phi =n = 0$. In this case we will use the full 
free energy in order to determine the phase diagram of the system.

\break
\subsection{Analytic Expressions in the Quartic Approximation}

We start by approximating the free energy with a quartic potential,
whose form we take to be
\label\LGpotdef
\eq
\beta [F - F(0) ] = \Bigl(a_\phi \, \phi^2 + a_n n^2\Bigr)
+  \nth{4} \Bigl( 
b_{\phi\phi} \phi^4 + 2\,b_{\phi n} \phi^2
n^2 + b_{nn} n^4  \Bigr) 
+ \cdots
\eeq
where the ellipsis denotes terms involving more than
four powers of $\phi$ or $n$. Using Eqs.~\FEresult\ 
and \evalresult\ and defining
\label\intdefs
\eq
	N_{n,m} = {1\over 2\hbar^{d-1}}\int {d^{d\!-\!1}\!k \; f^n\,g^m \over
	 (2 \pi)^d v_\ssf(k)}\,,\qquad \xi = \veps - \mu\,,
	\qquad T_\beta(\xi) = {\tanh(\beta \xi/2)\over\xi}\,,
\eeq
we find that
the quadratic coefficients are given by the following integrals:
\label\LGqda
\eqa
a_\phi &= a_\phi^0(\Lambda) 
-  N_{0,2}\int_{-\Lambda}^\Lambda d\veps \; T_\beta(\xi), \eol
\label\LGqdb
a_n  &= a_n^0(\Lambda) 
-  N_{2,0}\int_{-\Lambda}^\Lambda d\veps \; T_\beta(\xi)
\,{\xi\over\veps}\, , \eol
\eeq
where $a_\phi^0$ and $a_n^0$ are the ``bare'' values of these
quantities that appear in the original Hamiltonian,
eqs.~\Hamform--\Hamformthr.  The integral over $\veps$ diverges
logarithmically in the ultraviolet, forcing the introduction of the
cutoff scale, $\Lambda$. Physically this scale represents the energies
for which the linearization of the quasiparticle spectrum no longer
applies, or where the degrees of freedom of the Hamiltonian,
eqs.~\Hamform--\Hamformthr, are no longer appropriate, whichever is
lower.  If these approximations were exact, then $\Lambda$ would be the
actual maximum energy available on the lattice.  In any case, we can
consider $\Lambda$ to be of order the Fermi energy.  The divergent
dependence on $\Lambda$ as $\Lambda\to\infty$ can be absorbed into a
renormalization of $a_\phi^0$ and $a_n^0$.

The quartic coefficients are given similarly by the following
expressions, which since they are convergent allow us to take
the cutoff $\Lambda$ to infinity if we wish:
\label\LGqta
\eqa
b_{\phi\phi} &= 
- \, \; N_{0,4}\;\int_{-\Lambda}^\Lambda d\veps\; {T'_\beta(\xi)
\over \xi},\eol
\label\LGqtb
b_{\phi n} &= 
- \, \; N_{2,2}\;\int_{-\Lambda}^\Lambda d\veps\;{T'_\beta(\xi)
\over \veps},\eol
\label\LGqtc
b_{nn} &= 
- \, \; N_{4,0}\;\int_{-\Lambda}^\Lambda d\veps\;\left(
{T'_\beta(\xi)\over\veps} - {\mu\over\veps} \;
{\partial \over \partial \veps}\,\left({T_\beta(\xi)\over
\veps}\right)\right) . \eeol
\eeq
The approximation $\Lambda\to\infty$ is only valid
if $T$ and $\mu$ are much smaller than
the Fermi energy.

Because we work in the degenerate limit, near the FS, and at zero
field, the shape of the coefficient functions as $k$ varies over the FS
enters only as an overall normalization factor.  As expected from our
previous discussion, the $SO(5)$ symmetry is manisfestly present in the
limit $\mu= 0$ and $f^2/a^0_n = g^2/a^0_\phi$.

\subsection{Instability Towards Condensation}

Before turning to the numerical evaluation of these
integrals, it is instructive to first explore the 
physical implications of the renormalization itself. 
Because the sign of the loop contributions to $a_\phi$ and
$a_n$ is negative, these quantities
decrease as the temperature is decreased.  In the limit of
large $\beta\Lambda$, it is easy to see that the one-loop
contributions $\Delta a_\ssx = a_\ssx-a_\ssx^0$ have the
limiting behavior
\eq
\label\RGforqdcterms
	\Delta a_\phi \sim -2\,N_{0,2}\ln(\beta\Lambda);\qquad
	\Delta a_n \sim -2\,N_{2,0}\ln(\beta\Lambda).
\eeq
Even if
$a_\phi(T \sim \Lambda)$ and $a_n(T \sim \Lambda)$ are
initially chosen to be positive, so that $F$
is minimized by $\phi = n = 0$, eventually 
their decrease with decreasing $T$ can drive them negative, causing
an instability towards the development of nonzero
condenstates for $\phi$ or $n$. 

Eq.~\RGforqdcterms\ gives
the evolution of the \five-breaking combination of the electron
couplings:
\label\breakingrln
\eq
\beta\, {\partial \over \partial \beta}(a_\phi - a_n)  = 
  2 ( N_{2,0}- N_{0,2} )
 = 2 \int {d^{d\!-\!1}\!k \;\over (2 \pi)^d v_\ssf} \; (f^2 - g^2).
\eeq
This shows that any \five\ breaking present in the
original Hamiltonian due to $f^2\neq g^2$ will be enhanced in the
effective GL Hamiltonian by a potentially large logarithm of the form
$\ln(\beta\Lambda)$, where $\Lambda$ is of the order of the Fermi
energy.  For example,  suppose the AF
pairing is taken to be $s$-wave, $f=f_0$, whereas the SC
pairing is $d$-wave, corresponding to $g = g_0(\cos(a
k_x)-\cos(a k_y))$.  Here $f_0$ and $g_0$ are arbitrary constants.  In
general $N_{2,0}$ and $N_{0,2}$ differ from each other, and
although $g_0/f_0$ can be tuned to make the difference vanish, there seems
to be no symmetry principle that would require it to take this special
value.  Moreover, even if such a choice were made, it would still not
insure that the quartic couplings would have \five\ symmetry, \ie,
$N_{4,0}=N_{0,4}=N_{2,2}$.  In \S5 we will say more about the
extent of \five-breaking that arises from this choice of coupling
functions.

\subsection{Numerical Results near $\phi=n=0$}

Figures (1) and (3) present plots of the coefficients of the potential
as functions of the chemical potential, factoring out the integrals
over the FS.  In this section we describe these plots in more detail,
and use them to deduce the general shape of the phase diagram in the
temperature-doping plane.

Figure (1) plots the one-loop contribution to the quadratic
coefficients in the dimensionless form $(a_n-a_n^0)/N_{0,2}$ and
$(a_\phi-a_\phi^0)/N_{2,0}$, where $N_{n,m}$ are the FS integrals
defined in \intdefs.  At zero doping these functions are negative and
are increasing with temperature like 
$\ln(T/\Lambda)$, so it is always
possible to choose the bare coefficients $a_n^0$ or $a_\phi^0$ in such
a way that $a_n$ or $a_\phi$ is positive for high temperatures and
changes sign at some $T_c$, triggering the instability toward
condensation.   As the doping is increased, $a_n$ increases 
much more rapidly than $a_\phi$.  It follows that if
the AF phase is energetically preferable to the SC phase at $\mu=0$,
the system will be AF at zero doping, and then make the transition to
SC at some small doping, provided that $a_\phi$ is still negative at 
this value of $\mu$.

With applications to the high-$T_c$ cuprates in mind, we imagine
$\Lambda \sim 0.1 \eV \sim 10^3 K$ and so we adjust the bare coupling
$a_n^0$ to ensure that the AF transition (N\'eel) temperature,
$T_\ssn$, at $\mu=0$ is $1/10$ of the cutoff scale, $\beta_\ssn \Lambda
= 10$.  From fig.~1 it can be seen that at a temperature 
$T = T_\ssn/10$, for example,
we have $\beta \Lambda = 100$ and so the AF phase 
is quenched at a critical doping of $x_c \sim \mu_c/\Lambda = 10\,\%$. 
This is the doping at which $a_n(\mu_c)_{\beta\Lambda=100}$ becomes equal
to $a_n(0)_{\beta\Lambda=10}$.

\midinsert
\centerline{\epsfxsize=9.5cm\epsfbox{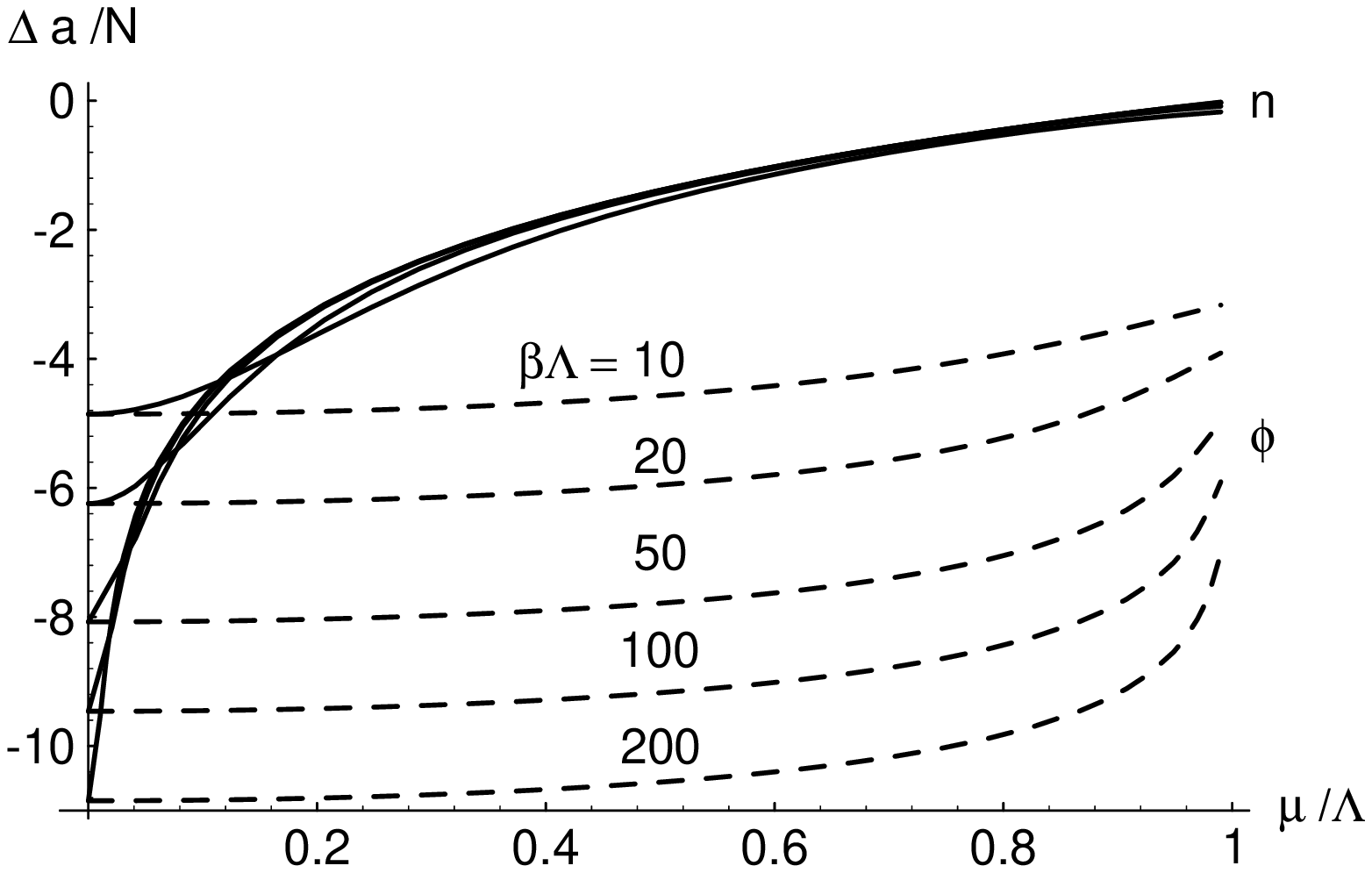}}
\centerline{\bf Figure 1}
\medskip\noindent
\vbox{\baselineskip=10pt \eightrm 
The coefficients $\ss (a_\phi-a^0_\phi)/N_{0,2}$ (dashed) and
$\ss (a_n-a^0_n)/N_{2,0}$ (solid) as a function of $\ss\mu/\Lambda$,
for $\ss\beta\Lambda = 10$, $\ss 20$, $\ss 50$, $\ss 100$ and $\ss 200$.}
\endinsert

The above example ignores the SC order parameter, which would be valid
if $a_\phi^0$ was taken to be much larger than $a_n^0$.  If on the
other hand \five\ symmetry was realized, so that $a_\phi^0=a_n^0$ and
$N_{0,2} = N_{2,0}$,  then at $\mu=0$ the ground state would be an
arbitrary admixture of AF and SC phases since both have the same
energy.  In this case, the system will always make a transition to the
pure SC phase at the critical doping $x_c$.  Since experiments indicate
that at $\mu=0$ the cuprates are pure AF, the SC phase must be
suppressed by having $a_\phi^0/N_{0,2} > a_n^0/N_{2,0}$, so that SC is
energetically disfavored until the critical doping is reached.  In this
situation, the phase diagram must have a form similar to figure 2.  Of
course, this prediction does not take account of quantum fluctuations within
the effective theory itself.  Such higher-loop effects are believed to be
especially important near the bicritical point, where the SC phase first
emerges \zhang.  They may have the effect of driving this point downward,
leaving a gap between the SC and AF phases at temperatures somewhat above
the existence of the bicritical point.

\topinsert
\centerline{\epsfxsize=7.5cm\epsfbox{fig2.eps}}
\vskip-2in
\centerline{\bf Figure 2}
\medskip\noindent
\vbox{\baselineskip=10pt \eightrm 
Schematic representation of the phase diagram for the present
Hamiltonian, with $\ss SO(5)$ symmetry broken to favor the
AF condensate.  Quantum effects not considered here may be able to
drive the bicritical point downward.}
\endinsert

To determine exactly how much \five\ breaking is needed in the
quadratic terms in order to suppress SC at zero doping, one must
compare the local minima of the potential along the $\phi=0$ and $n=0$
axes and insure that the minimum with $n\neq 0$ is the deeper of the
two.  This issue can usually be settled by looking at the higher order
terms in the potential.  To this end we have computed the quartic
coefficients in the dimensionless forms $ b_{\phi\phi}T^2 / N_{0,4}$,
$b_{nn}T^2/ N_{4,0}$ and $b_{n\phi}T^2/ N_{2,2}$, and plotted them in
fig.~3.

The fact that $b_{\phi\phi}$ is only weakly dependent on the doping can be
see by observing that when $\Lambda\to\infty$, $\mu$ can be
completely removed by shifting the integration variable from
$\veps$ to $\xi$. The low temperature form of the quartic 
couplings can be further understood analytically through the identities
\eqa
\label\deltaa
	\lim_{\beta\to \infty} 
       {T'_\beta(\xi)\over \beta^2\, \xi} &= 
	-{7\zeta(3)\over 2\pi^2}\,\delta(\xi)\eol
\label\deltab
	\lim_{\beta\to \infty} {1 \over \beta^2} \;
     \left[ {T'_\beta(\xi)\mu^2\over \xi\veps^2}
	-{\mu T_\beta(\xi)\over\veps^3} \right] &=
	-{7\zeta(3)\over 2\pi^2}\,\delta(\xi)\eeol
\eeq
Eq.~\deltab\ is valid so long as it is combined with a function that
is nonsingular at $\veps=0$, in which case its own $1/\veps$ singularity
can be integrated over using the principal value.  Both of these
representations of the delta function can break down if multiplied by
singular functions of $\xi$, which is precisely what happens at $\mu=0$,
but for $\mu\gg 1/\beta$ they can be used in conjunction with
\LGqta--\LGqtc\ to deduce that $b_{nn} = b_{n\phi} =0$ and
$b_{\phi\phi}/N_{4,0} = 7\zeta(3)/2\pi^2$, in the low temperature limit. 

At zero doping, all three coefficients have essentially the same value,
and from the preceding paragraph we understand why $b_{\phi\phi}$ remains
nearly constant up to very large doping, whereas $b_{nn}$ and $b_{n\phi}$
drop quite sharply.  In fact, they reach {\it negative} values for
intermediate dopings.  For these dopings, the truncation to fourth order
in the fields gives a potential which is unbounded from below.  Thus the
expansion of the potential cannot be used to determine the minimum in the
$n$ direction, and our goal of finding the absolute minimum cannot be
attained by expanding only to quartic order.  We must use the full
functional form of the free energy for this purpose. 

\midinsert
\centerline{\epsfxsize=12.5cm\epsfbox{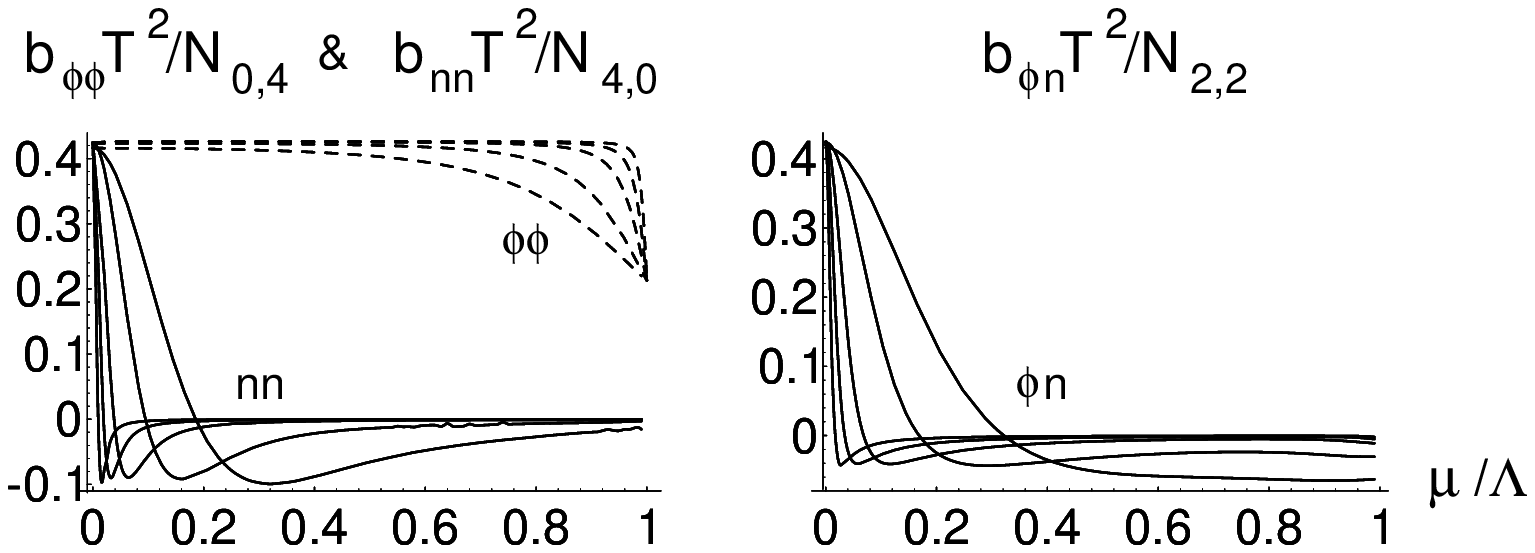}}
\centerline{\bf Figure 3}
\medskip\noindent
\vbox{\baselineskip=10pt \eightrm 
The coefficients $\ss b_{\phi\phi}T^2/ N_{0,4}$,
$\ss  b_{nn}T^2/ N_{4,0}$   and
$\ss b_{n\phi}T^2/ N_{2,2}$
as a function of $\ss\mu/\Lambda$,
for $\ss\Lambda/T = 10$, $\ss 20$, $\ss 50$, $\ss 100$ and $\ss 200$.
The curves approach an asymptotic limit as $\ss\Lambda/T\to\infty$.}
\bigskip
\endinsert

\subsection{Numerical Results for Arbitrary $\phi$ and $n$}

In computing the full expression for $F_1$ in eq.~\FEresult, we can no
longer factorize the momentum integral into a product of the form $\int
d^{d\!-\!1}\!k\, G(k) \int d\veps H(\veps)$ because the integrand is not a
simple polynomial in $f_p$ and $g_p$.  However we can still treat it
similarly to our previous computations if we consider the special case
where $f^2$ and $g^2$ are constant.  We do not expect this assumption to
radically change the features of the more general case, since the
$k$-dependence of $f$ and $g$ has played no essential role in the
preceding discussion.  Constant $g^2$ still describes $d$-wave SC and is
consistent with \five\ symmetry, so long as $g$ changes sign across the
nesting surface, {\it e.g.} $g_p=$ sign$(\veps_p)$.

%Moreover the conditions $f_p=1$ and $g_p =
%$ sign$(\veps_p)$ have many of the essential features of the real cuprate
%systems, since the former correctly describes $s$-wave AF pairing
%while the latter is similar to the lattice expression for $d$-wave
%pairing, $g_p = \cos(p_x a)- \cos(p_y a)$, in that both expressions
%are odd under translations by the nesting vector, $g_{p+Q} = -g_p$.

With this assumption, the free energy reduces to a single integral over
the energy.  We have evaluated it numerically for the relevant ranges
of doping, temperature, and order parameters.  The results are most
conveniently expressed in terms of the following variables, for
which the cutoff, $\Lambda$, is scaled out:
\label\hatdefs
\eqa
	\hat\phi &= g\phi/\Lambda;\quad \hat n = fn/\Lambda;\quad
	x = \mu/\Lambda; \eol
	\hat a_\phi &= a^0_\phi/(4 N_{0,0} g^2);\quad
	\hat a_n = a^0_n/(4 N_{0,0} f^2);\quad \hat\beta = \beta\Lambda/2\eol
	\hat\veps &= \veps/\Lambda;\quad\hat\lambda_\pm =
	 ( (\sqrt{\hat\veps^2+ \hat n^2}\pm x)^2 + \hat\phi^2 )^{1/2};\eol
	\Delta \hat F &= {1\over\hat\beta} \int_0^1 d\hat\veps 
  \ln\left({\cosh(\hat\beta\hat\lambda_+)\cosh(\hat\beta\hat\lambda_-)\over
  \cosh(\hat\beta(\hat\veps+x))\cosh(\hat\beta(\hat\veps-x))}\right)\eeol
\eeq
In the last line we subtracted a constant to make $\Delta\hat F$ vanish
at $\phi=n=0$.  Then the full effective potential can be written in terms of 
\label\hatF
\eq
	\hat F \equiv F/(4N_{0,0}\Lambda^2) = 
	\hat a_\phi \hat\phi^2 + 
	\hat a_n \hat n^2 + 
	\Delta\hat F(\hat\phi,\hat n,\hat\beta,x).
\eeq

\topinsert
\bigskip\bigskip
\centerline{\epsfxsize=12cm\epsfbox{fig4.eps}}
\vskip-0.8in
\centerline{\bf Figure 4}
\medskip\noindent
\vbox{\baselineskip=10pt \eightrm 
The rescaled effective potential $\ss \hat F$ for six values of
the doping parameter, $\ss x$ (horizontal rows) and along three
directions in the $\ss \phi$-$\ss n$ plane (vertical columns).  
$\ss \hat F$ is plotted as a function of $\ss\hat\phi^2$,
$(\ss\hat\phi^2+\hat n^2)/2$, and $\ss\hat n^2$, in the three
respective columns.  Each graph is shown for the range of temperatures
given by $\ss \beta\Lambda = 4,\ 10,\ 20,$ and $\ss 200$, with
$\ss \hat F$ always increasing with temperature.}
\bigskip \endinsert

In figure 4 we show $\hat F$ for six values of the doping parameter, $x
= 0$, 0.2, 0.4, 0.6, 0.8 and 1 (horizontal rows), and along three
directions in the $\phi$-$n$ plane (vertical columns).  $\hat
F$ is plotted as a function of $\hat\phi^2$, $(\hat\phi^2+\hat n^2)/2$,
and $\hat n^2$, in the three respective columns.  Each graph is shown
for the range of temperatures given by $ \beta\Lambda = 4,\ 10,\ 20,$
and $200$, with $\hat F$ always increasing with temperature.  We
have chosen $\hat a_\phi = 2$ and $\hat a_\phi = 1.6$, hence
\five-breaking at the 20\% level, in order to illustrate the
phenomenologically desired dominance of the AF phase at low doping.  At
$x=0$, the lowest minimum clearly occurs for $n\neq 0$.  At $x = 0.2$,
the minima become degenerate and the transition to the SC phase
begins.  Further increasing the doping quickly removes the AF
instability, and gradually quenches the SC phase as well, depending on
the temperature.  This is in complete agreement with the picture
obtained from our analysis of the quadratic coefficients and confirms
the phase diagram shown in fig.~2.

\break
\subsection{Terms with Derivatives}
So far we have presented the one-loop free energy only for spatially
constant order parameters.  If $\phi(x)$ and $n(x)$ are not constant,
there will be additional terms depending on their spatial derivatives.
Of these, the most interesting are the $(\grad\phi)^2$ and $(\grad
n)^2$ terms, since they are needed to compute quantum corrections
coming from within the effective long-wavelength theory (two-loop
and higher corrections).

By calculating the relevant Feynman diagrams, it is straightforward to
show that the quadratic terms in $F_1$ for $\phi_q$ and $n_q$ with an
arbitrary momentum $q$ can be obtained from our previous formulas
\LGqda--\LGqdb\ by making the replacements
\eqa
	T_\beta(\xi) &\to {\tanh(\beta\xi_{p+q/2}/2)+\tanh(\beta\xi_{p-q/2}/2)
	\over \xi_{p+q/2} + \xi_{p-q/2}},\eol
	f(p,q) &\to f(p-q/2,q),\qquad g(p,q) \to g(p-q/2,q),\eeol
\eeq
where $\xi_p = \veps_p-\mu$ and $p$ is the loop momentum to be
integrated over. Expanding in the external momentum $\vec q$ gives the
expansion in derivatives of the fields.  From this prescription it is
clear that \five\ symmetry will be realized in the derivative terms so
long as $f^2(p,q) = g^2(p,q)$ and $\mu=0$, since the symmetry is then
manifest in the energy eigenvalues $\lambda_\pm$ that appear in
\evalresult.

The resulting integral over $p$ depends on the exact form of the
dispersion relation $\veps_p$ and the coupling functions $f_p$, $g_p$,
so we will defer further evaluation until the next section where we
consider a specific choice of these functions.

\section{Exact Results in a Specific Model}

Our analysis so far has relied upon the approximation that the most
important contributions to the effective potential are coming from
electronic excitations near the Fermi surface.  This approximation
presupposes the Fermi energy to be much greater than the temperature
and the chemical potential.  On the other hand this hierarchy of scales
may only be a factor of 10 in the actual cuprate systems we are
interested in.  It is therefore interesting to corroborate our
conclusions without assuming that $F_1$ is dominated by the
contributions from near the FS.  In this section we will exactly
evaluate the first few terms in the small-field expansion of $F$ using
a semi-realistic lattice dispersion relation and pairing functions,
which have been previously suggested in the literature
[1,5,9].  It will be shown that the results are
in good qualitative agreement with those based on the RG approach.

\subsection{Method of Solution}

For our specific model, we take the lattice dispersion relation
\dispex\ of the extended Hubbard model and choose 
\eq 
\label\fgdefs
 f(p,q) =1, \qquad g(p-q/2,q) = (\cos(a p_x) - \cos(a p_y))/2 
\eeq 
as would be appropriate for $s$-wave AF pairing and $d$-wave SC
pairing.  The factor of 2 in the definition of $g$ is chosen such as
to maximize the \five\ symmetry near the undoped FS, as is shown
below, and the $q$ dependence is chosen to simplify the derivative
terms to be computed at the end of this section.  
We further assume that the system is two dimensional, $d=2$.

The choice \fgdefs\ is motivated by two considerations: (1) we wish to
examine to what extent \five\ symmetry survives when $f^2\neq g^2$,
since it has been suggested that \five\ can be a good approximate
symmetry in the long-wavelength limit, even if it is not exact in the
fundamental Hamiltonian; and (2) one would like to see how much the
general features of the phase diagram depend on having exact
\five\ symmetry in the Hamiltonian.  We will show that making $f^2\neq
g^2$ breaks the symmetry significantly due to fluctuations far from the
Fermi surface, even though those close to the FS give contributions to
the free energy which are approximately \five-symmetric.  Nevertheless,
the phase diagram is not affected in a qualitative way, and should
still have the shape shown in figure 2.

The effective potential at $q=0$
is still given by the general expression \FEresult, but now
the momentum integrals range over the Brillouin zone, $-\pi/a < p_x,p_y
< \pi/a$, and there is no need to introduce any cutoff because the
energy is explicitly bounded by $-4t < \veps < 4t$.   As before, we
expand $F$ to quadratic and quartic order in the fields.  Defining
$\xi_p =  \veps_p - \mu$, this gives
\eq
	F_1 = -\int{d^{\,2}p\over(2\pi)^2}\Biggl(  2
	T_\beta(\xi_p)\left(|g_p\phi|^2 + n^2{\xi_p\over\veps_p}\right)
	+ {T'_\beta(\xi_p)\over 2\xi_p} 
	\left(|g_p\phi|^2 + n^2{\xi_p\over\veps_p}\right)^2
	+ {\mu\over 2\veps^3_p}T_\beta(\xi_p) n^4 \Biggr). 
\eeq

Next one would like to reexpress the momentum integrals as an integral
over the energy times an integral over the Fermi surface.  This can be
accomplished using the identity
\eq
	\int d^{\,2} p\, g_p^{m} = {4\over a^2} \int_{-4t}^{4t} d\veps
	\int_{-1}^1 \int_{-1}^1 
	{du\,dv\,(u-v)^{m}2^{-m}\over\sqrt{1-u^2}\sqrt{1-v^2}}
	\,\delta(\veps + 2t(u+v)),
\eeq
where $u = \cos(a p_x)$ and $v = \cos(a p_y)$.  The integrals over $u$ and
$v$ can be done exactly, leaving an energy integral times a density of
states function,
\eq
	\int d^{\,2} p\, g_p^{m} = {4\over a^2 t}\int_{-4t}^{4t} d\veps
	\,{\cal N}_{m}(\veps/4t).
\eeq
The density of states ${\cal N}_m$ is given in terms of complete elliptic
functions $K$ and $E$; \foot\ellfns{as they are defined in all standard 
mathematical references except for Abramovitz and Stegun and 
Mathematica, which use a nonstandard convention.  See for example \Efns}
defining $\hat\veps = \veps/4t$ and $\alpha = 
(1+|\hat\veps|)/(1-|\hat\veps|)$, the first few are 
\eqa
	{\cal N}_{0}(\hat\veps) &= {1\over 1+|\hat\veps|}\, K(1/\alpha)\eol
	{\cal N}_{2}(\hat\veps) &= (1+|\hat\veps|)
	\bigl( K(1/\alpha)-E(1/\alpha)\bigr)\eol
	{\cal N}_{4}(\hat\veps) &= {1\over 3}\,\alpha (1-|\hat\veps|)^3
	\bigl( (2\alpha^2+1)K(1/\alpha)-2(\alpha^2+1)E(1/\alpha)\bigr)\eeol
\eeq
These are shown in fig.~5.  Because of our choice of normalization for
$g_p$ in \fgdefs, they all have the same asymptotic behavior near the
undoped FS, $\veps = 0$: ${\cal N}_{m} \sim \ln(1/|z|)$, independent of
$m$.  However they quickly diverge from each other for energies above
or below the FS.  This is interesting from the point of view of the
\five\ symmetry, because it shows that even though $g^2\neq 1$, as is
necessary to have exact \five, close to the FS the symmetry is
nevertheless approximately realized.  However the logarithmic
singularity near $\veps=0$ in ${\cal N}_m$ is too weak to compensate
for the \five-breaking differences between ${\cal N}_0$ and ${\cal
N}_2$ away from the FS, as we will presently show.

\midinsert
\bigskip\bigskip
\centerline{\epsfxsize=8cm\epsfbox{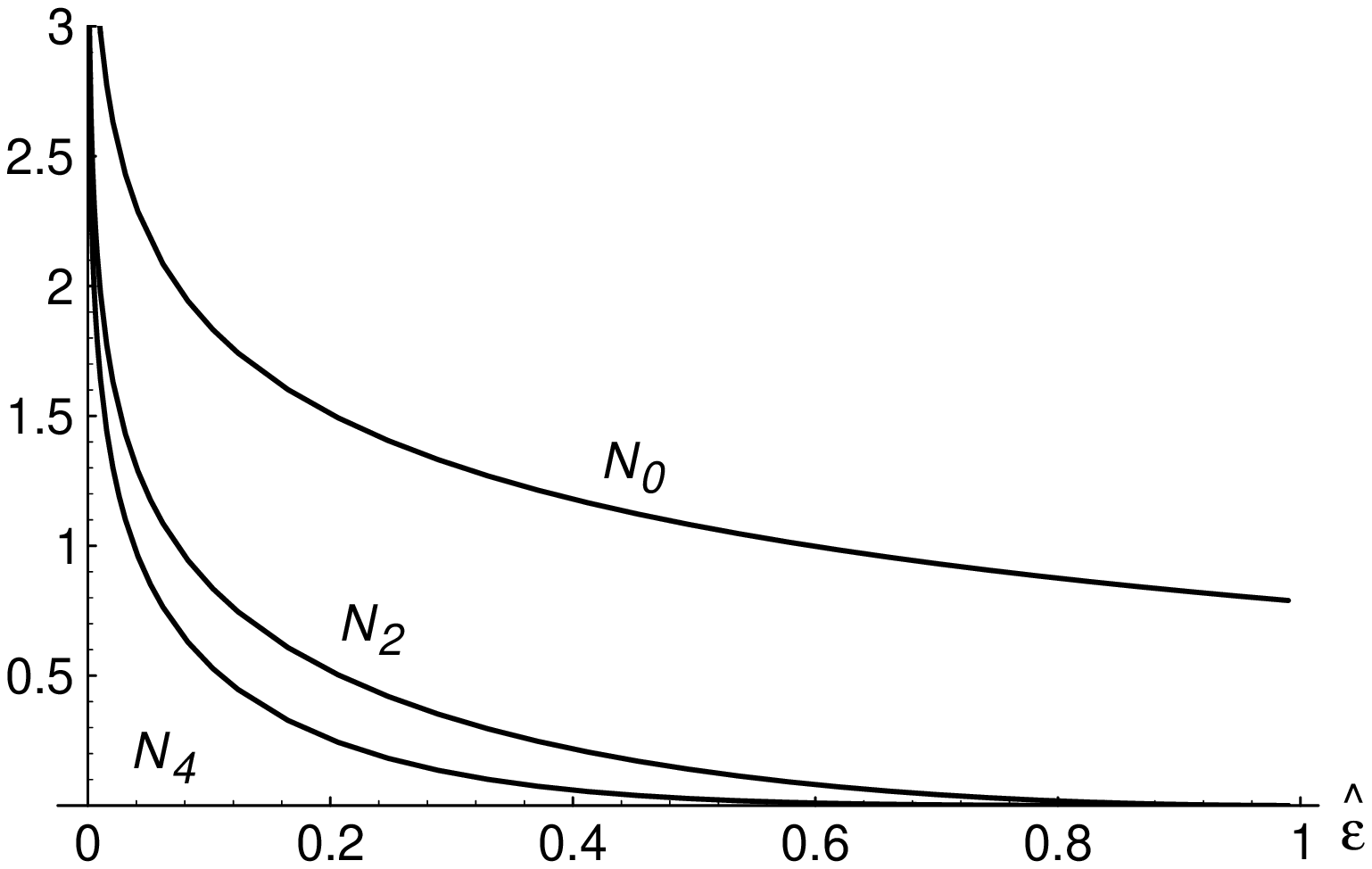}}
\centerline{\bf Figure 5}
\medskip\noindent
\vbox{\baselineskip=10pt \eightrm 
The density of states ${\cal N}_0$, ${\cal N}_2$ and ${\cal N}_4$
as a function of $\hat\veps = \veps/4t$.}
\bigskip \endinsert

\subsection{Expansion near $\phi=n=0$}

Now the expressions for the coefficient functions in the small-field
expansion of $F_1$ can be written in a way which looks very similar
to our previous results \LGqda--\LGqtc.  Let us first define
\eq
\kappa = {4\over \hbar^2 (2\pi)^2a^2 t} = {2m\over(\pi\hbar)^2}, 
\eeq
where $a$ is the lattice spacing, and $m$ is the quasiparticle mass in
the vicinity of the FS, where the dispersion relation can be
approximated as $\veps = p^2/2m$.  We then obtain
\label\LGqdaa
\eqa
a_\phi &= a_\phi^0
- \kappa\int_{-4t}^{4t} d\veps \; T_\beta(\xi)\, 
{\cal N}_{2}(\veps/4t)\eol
\label\LGqdbb
a_n  &= a_n^0
- \kappa\int_{-4t}^{4t} d\veps \; T_\beta(\xi)\,
{\xi\over\veps}\,{\cal N}_{0}(\veps/4t) , \eol
\label\LGqtaa
b_{\phi\phi} &= 
- \, \kappa\;\int_{-4t}^{4t} d\veps\; {T'_\beta(\xi)
\over \xi}\,{\cal N}_{4}(\veps/4t) ,\eol
\label\LGqtbb
b_{\phi n} &= 
- \, \kappa\;\int_{-4t}^{4t} d\veps\;{T'_\beta(\xi)
\over \veps}\,{\cal N}_{2}(\veps/4t),\eol
\label\LGqtcc
b_{nn} &= 
- \, \kappa\;\int_{-4t}^{4t} d\veps\;\left(
{T'_\beta(\xi)\over\veps} - {\mu\over\veps}
\; {\partial \over \partial \veps} \, 
\left({T_\beta(\xi)\over
\veps}\right)\right) {\cal N}_{0}(\veps/4t). \eeol
\eeq

The dependence of these coefficients on doping and temperature is shown
in Figures 6 and 7, which can be compared to Figures 1 and 3 from our
model-independent analysis.  In Figure 6, which shows the 1-loop
contributions to the quadratic coefficients, one sees that our previous
result for $a_n$ is in good agreement with the exact computation in the
present model.  This is not suprising because $f_p$ is treated as a
constant in both cases, and the only difference between the two is the
lattice dispersion relation in the present computation versus the
approximation $\veps \approx v_\ssf \ell$ in \S4.  However the
difference between the two approaches is quite apparent in $a_\phi$,
where $g_p$ is relevant:  $a_\phi$ increases with doping faster in the
exact computation than in the approximation of factoring out the FS
integration.  Nevertheless the important qualitative feature needed to
obtain the phase diagram of Figure 2 persists:  $a_\phi$ increases more
slowly as a function of doping than does $a_n$.

\midinsert
\bigskip\bigskip
\centerline{\epsfxsize=12.5cm\epsfbox{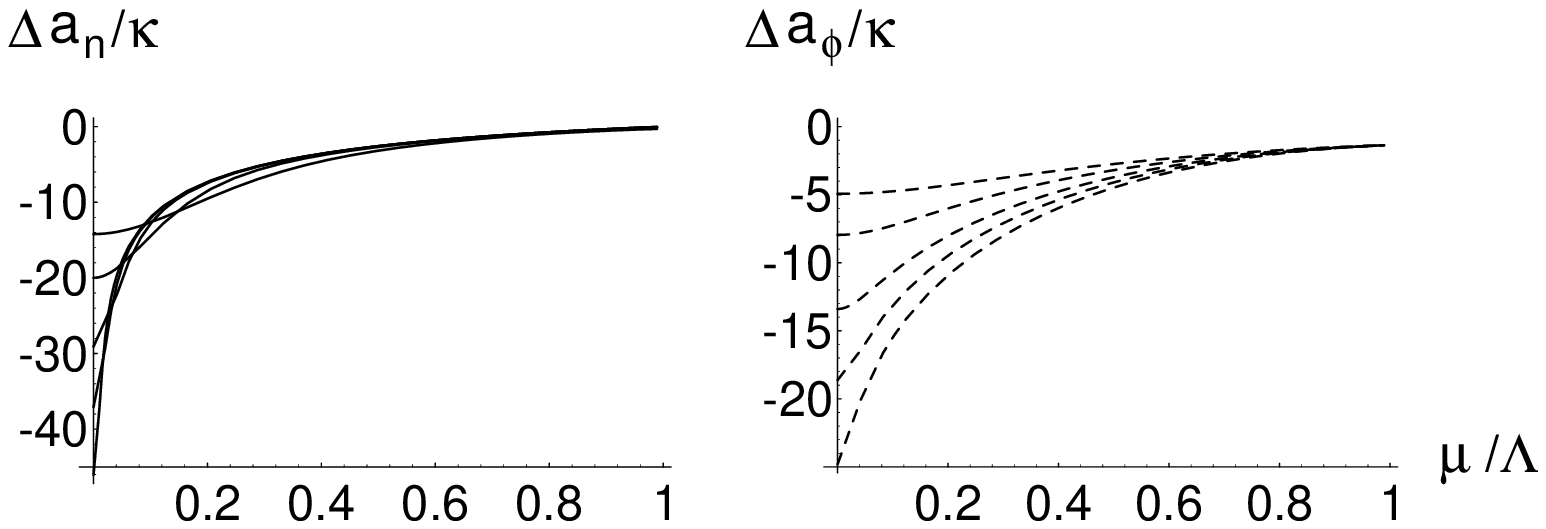}}
\centerline{\bf Figure 6}
\medskip\noindent
\vbox{\baselineskip=10pt \eightrm 
The exactly computed coefficients 
$\ss (a_n-a^0_n)/\kappa$ (solid) and
$\ss (a_\phi-a^0_\phi)/\kappa$ (dashed) 
as a function of $\ss\mu/\Lambda$,
for the same values of $\ss\beta\Lambda$ as in Figures 1 and 3.}
\bigskip 
\centerline{\epsfxsize=12.5cm\epsfbox{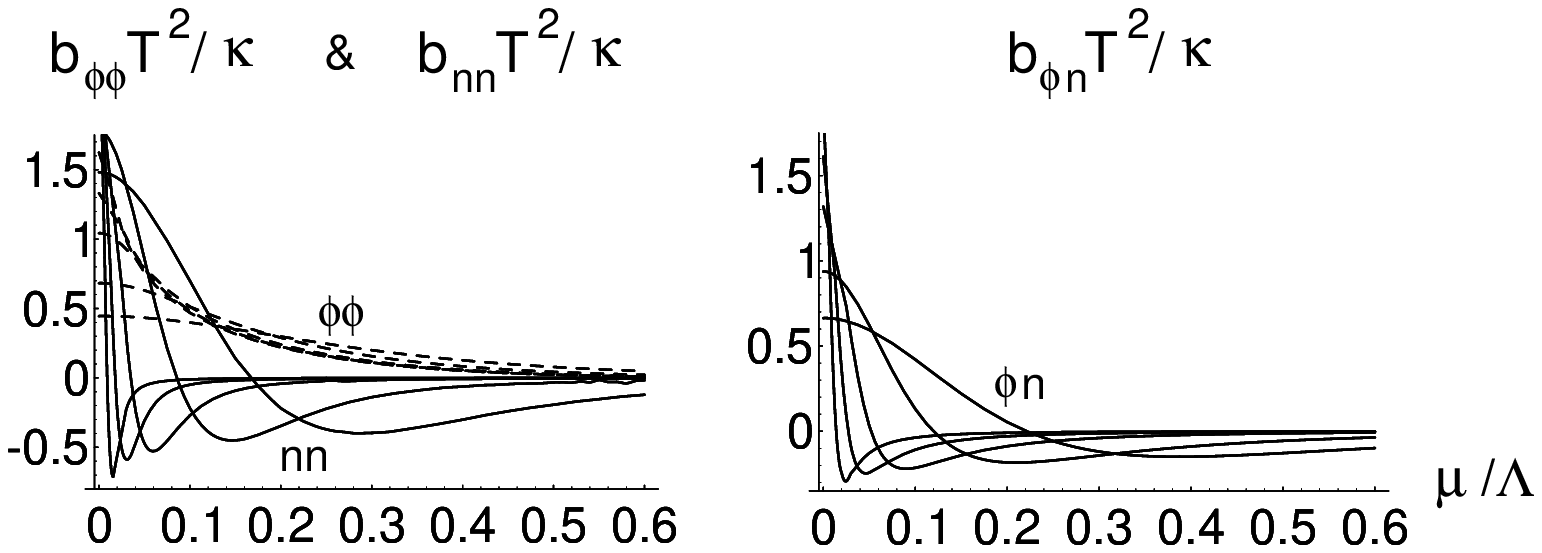}}
\centerline{\bf Figure 7}
\medskip\noindent
\vbox{\baselineskip=10pt \eightrm 
The exactly computed coefficients 
$\ss b_{nn} T^2/\kappa$, $\ss b_{n\phi} T^2/\kappa$ and $\ss b_{\phi\phi} 
T^2/\kappa$ as a function of $\ss\mu/\Lambda$,
for the same values of $\ss\beta\Lambda$ as in Figures 1, 3 and 6.}
\bigskip \endinsert

Similar remarks apply in comparing the quartic coefficients in the two
treatments.  The difference in the dispersion relation causes a small
change in the $T$- and $\mu$-dependence of $b_{nn}$, the coefficient
that does not depend at all on $g_p$.  Nevertheless in both analyses,
$b_{nn}$ and $b_{n\phi}$ have the same behavior of initially falling
very sharply from positive to negative values as the chemical potential
$\mu$ is increased from $\mu=0$, and gradually becoming zero as $\mu\to
4t$.  $b_{\phi\phi}$ falls much faster with $x$ in the exact
computation than in the previous one, but still more gradually than
$b_{nn}$ or $b_{n\phi}$.  Again, this can be understood analytically in
the low-temperature limit using \deltaa--\deltab, which gives $b_{nn} =
b_{n\phi} = 0$ and $b_{\phi\phi} = (7\zeta(3)\beta^2
m/\pi^4\hbar^2){\cal N}_4(\mu/4t)$ for $\mu\gg 1/\beta$. The
qualitative agreement between the two approaches gives us confidence
that the shape of the phase diagram in Figure 2 is indeed a robust
prediction of the model, in the one-loop approximation.

\break
\subsection{\five-breaking due to $g^2_p\ne f^2_p$}

One interesting issue we can explore within the exact computation is
the extent to which \five\ symmetry is broken by the assumption of
$d$-wave SC pairing versus $s$-wave AF pairing.  In Fig.~6 one sees
that, despite our tuning the normalization of $g_p$ so as to preserve
\five\ near the FS, $|\Delta a_n| > |\Delta a_\phi|$ even at zero
doping.  We could have tuned $g_p$ such as to force $\Delta a_n =
\Delta a_\phi$ at some temperature $T_0$, but the equality would hold
only for $T=T_0$.  In fact we find that the ratio $\Delta
a_\phi/\Delta a_n$ scales linearly with $\ln(\beta/4t)$:
\eq
	\left({\Delta a_\phi\over\Delta a_n}\right)_{\mu=0}
	 \cong 0.20 + 0.064\,\ln(\beta/4t).
\eeq
For comparison, the breaking in the quartic couplings can be similarly
parameterized in terms of the ratios
\eq
   \left({b_{\phi\phi}\over\ b_{nn}}\right)_{\mu=0} 
   \cong  0.082 + 0.10\,\ln(\beta/4t);\qquad
   \left({b_{n\phi}\over\ b_{nn}}\right)_{\mu=0} 
   \cong  0.27 + 0.084\,\ln(\beta/4t).
\eeq

\subsection{Derivative Terms in the Free Energy}
Using the above techniques and the procedure discussed in \S4.5, we can
evaluate the lowest derivative terms in the quadratic part of $F_1$,
\eq
	F_1 = c_\phi (\grad\phi)^2 + c_n (\grad n)^2.
\eeq
Defining $\kappa' = \beta^2 t/(4\pi\hbar)^2$, $\hat\veps = \veps/4t$,
$\hat\mu = \mu/4t$,  and going to the limit
of large $\beta t$, we can write the coefficients as
\eqa
\label\qddera
 c_\phi &\cong \kappa' \int_{-4t}^{4t} d\veps \; {T_\beta(\xi)\over
	\cosh^2(\beta\xi/2)}\left( 
	\left(1-\hat\veps^2\right){\cal N}_2(\hat\veps) -
	\left(1-|\hat\veps|\right)^2{\cal N}_4(\hat\veps)\right)\eol
\label\qdderb
&\cong {14\zeta(3)\kappa'\over\pi^2} \left(
	\left(1-\hat\mu^2\right){\cal N}_0(\hat\mu) -
	\left(1-|\hat\mu|\right)^2{\cal N}_2(\hat\mu)\right);\eol
\label\qdderc
c_n &\cong \kappa' \int_{-4t}^{4t} d\veps \; {T_\beta(\xi)\over
	\cosh^2(\beta\xi/2)}\left({\xi\over\veps}\right)\left(
	\left(1-\hat\veps^2\right){\cal N}_0(\hat\veps) -
	\left(1-|\hat\veps|\right)^2{\cal N}_2(\hat\veps)\right)\eol
\label\qdderd
&\sim  O(1/\beta t)\,c_\phi \eeol
\eeq
To obtain \qdderb\ we used the identity \deltaa, which is however only
valid for $\mu\gg 1/\beta$.  For $\mu\lsim 1/\beta$, the logarithmic
singularities in ${\cal N}_0$ and ${\cal N}_2$ are integrated over to
obtain result that is finite as $\mu\to 0$.  The identity
\deltaa\ gives a vanishing result when applied to \qdderc, meaning that
$c_n$ is of order $1/\beta t$ compared to $c_\phi$, and thus $c_n$ must
be computed numerically to find the leading behavior in the large
$\beta t$ limit.  Again, these statements hold only for $\mu\gg
1/\beta$ due to the singular behavior of the integrand at $\veps=0$.
For $\mu\lsim 1/\beta$, we expect that $c_\phi\sim c_n$, just as was
found for the coefficients of the quartic terms.

These derivative terms will not be of further use in the present 
preliminary study, but may be useful in computing the higher order
quantum corrections which we leave for future investigation.

\section{Conclusions}

In summary, our calculations lead us to the following 
conclusions:

\topic{Microscopic $SO(5)$ Invariance}
We find, by construction, that microscopic models can indeed give rise
to \five\ invariance which relates the $d$-wave superconducting and
antiferromagnetic phases. This symmetry can emerge accidentally at long
wavelengths, and so be relevant for the effective potential which
determines which phase is energetically preferred (in a homogeneous
system). Being a long-wavelength effect, the symmetry can be easily
hidden in treatments which include all scales together.

\topic{Features of the Phase Diagram}
Using our model we have computed the Landau-Ginzburg free energy whose
minimization determines the phase diagram in the temperature-doping
plane. We find that the coefficients of the Landau-Ginzburg potential
have the right qualitative dependence on temperature and chemical
potential to allow for antiferromagnetism at half filling, but which is
destroyed as the system is doped. It can also describe a superconductor
which is favoured only for a range of nonzero dopings.

\topic{Significance of the \five\ Symmetry} 
The class of models we considered encompasses \five\ symmetry as a
special case among the most general possible coupling functions.  We
find that if the symmetry is broken in the microscopic theory, there is
no special tendency for it to be restored or further broken in the
effective long-wavelength theory.  The phase diagram is only of the
desired form if there is a certain level of \five-breaking, whose
fundamental origin is unknown.  In the specific model considered in
\S5, the symmetry appears to be badly broken, since $f^2_p$ and $g^2_p$
are quite different functions of momentum, yet by tuning their relative
sizes it appears possible to obtain the correct phase diagram.  At
present it is therefore not yet clear whether nature demands an
approximate \five\ symmetry as the correct way to understand the
relation between the AF and SC phases in cuprates, or whether
\five-symmetric models simply fall into the same universality class as
the correct theory.

\endtopic

\bigskip
\centerline{\bf Acknowledgments}
\bigskip

We thank the organizers and participants of the
Benasque Centre for Physics, for
providing such a pleasant and stimulating setting for part
of this work. We thank P.K. Panigrahi for useful discussions,
and C.A. L\"utken for collaborations in early stages of this
work. Our research was partially funded 
by the N.S.E.R.C.\ of Canada and by the Fonds F.C.A.R. du Qu\'ebec.

%{\baselineskip 15pt
\listrefs
%}
%\figurecaptions

\bye

%% file: macros.tex
% ------------------------------------------------------------------
% Size and shape
% ------------------------------------------------------------------

\font\titlefont = cmr10 scaled\magstep 4
 2
\font\sectionfont = cmr10
\font\littlefont = cmr5 % for equation names in draftmode
\font\eightrm = cmr8 

\def\ss{\scriptstyle} 
\def\sss{\scriptscriptstyle} 

\newcount\tcflag
\tcflag = 0  % this flag indicates whether harvmac is to be used
 %turn off 12 point definition if run with harvmac

\ifnum\tcflag = 0 \magnification = 1200 \fi  % 12 point

%\global\hsize = 5in % size of text
%\global\lmargin = 0.125in% 
\global\baselineskip = 1.2\baselineskip % line skip
\global\parskip = 4pt plus 0.3pt % paragraph skip
\global\abovedisplayskip = 18pt plus3pt minus9pt
\global\belowdisplayskip = 18pt plus3pt minus9pt
\global\abovedisplayshortskip = 6pt plus3pt
\global\belowdisplayshortskip = 6pt plus3pt

\def\barsoff{\overfullrule=0pt}

% ------------------------------------------------------------------
% Draft mode stuff
% ------------------------------------------------------------------

\def\endignore{}
\def\ignore #1\endignore{} % use to "comment out" text

\newcount\dflag% draft mode flag
\dflag = 0% initialize

% Time commands ---------------------------------------------------

\def\monthname{\ifcase\month % for month numbers
\or January \or February \or March \or April \or May \or June%
\or July \or August \or September \or October \or November %
\or December % monthname 
\fi}

\newcount\dummy
\newcount\minute  % defines counters for the timestamp
\newcount\hour
\newcount\localtime
\newcount\localday
\localtime = \time
\localday = \day

\def\advanceclock#1#2{ % advances clock to give local time
\dummy = #1
\multiply\dummy by 60
\advance\dummy by #2
\advance\localtime by \dummy
\ifnum\localtime > 1440 % advances day if clock is advanced past midnight
\advance\localtime by -1440
\advance\localday by 1
\fi}

\def\settime{{\dummy = \localtime %
\divide\dummy by 60%
\hour = \dummy % hour
\minute = \localtime%
\multiply\dummy by 60%
\advance\minute by -\dummy % minutes
\ifnum\minute < 10 
\xdef\spacer{0} % leading 0 for minutes
\else \xdef\spacer{} 
\fi %
\ifnum\hour < 12 
\xdef\ampm{a.m.} % before noon
\else 
\xdef\ampm{p.m.} % after noon
\advance\hour by -12 %
\fi %
\ifnum\hour = 0 \hour = 12 \fi % make midnight, noon = 12
\xdef\timestring{\number\hour : \spacer \number\minute%
\thinspace \ampm}}}

% Draft mode commands --------------------------------------------

% ------------------------------------------------------------------
% Headers
% ------------------------------------------------------------------

\def\endtitle{}
\def\title#1\endtitle{\vskip.5in\titlefont
\global\baselineskip = 2\baselineskip % set line skip
#1\vskip.4in% title
\baselineskip = 0.5\baselineskip\rm}
 
\def\endauthors{}
\def\authors#1\endauthors{#1}

\def\endabstract{}
\def\abstract#1\endabstract{\vskip .3in%
\centerline{\sectionfont\bf Abstract}%
\vskip .1in
\noindent#1}

\def\nopageonenumber{\footline={\ifnum\pageno<2\hfil\else
\hss\tenrm\folio\hss\fi}}  % turns off page number on title page

\newcount\nsection % section counter
\newcount\nsubsection % subsection counter

% start new section
\def\section#1{\global\advance\nsection by 1% increment section number
\nsubsection=0
% section title
\bigskip\noindent\centerline{\sectionfont \bf \number\nsection.\ #1}
\bigskip\rm\nobreak}% back to normal

% start new subsection
\def\subsection#1{\global\advance\nsubsection by 1% increment subsection number
% subsection title
\bigskip\noindent\sectionfont \sl \number\nsection.\number\nsubsection)\
#1\bigskip\rm\nobreak}% back to normal

% unnumbered itemized topics 
\def\topic #1{{\medskip\noindent $\bullet$ \it #1:}} 
\def\endtopic{\medskip}

% start new appendix
\def\appendix#1#2{\bigskip\noindent%
\centerline{\sectionfont \bf Appendix #1.\ #2} % appendix title
\bigskip\rm\nobreak} % back to normal

% ------------------------------------------------------------------
% References
% ------------------------------------------------------------------

\newcount\nref % create counter for references
\global\nref = 1 % initialize it

% this is just in case there are no references at all but \listrefs
% is nevertheless used
\def\therefs{} 

 % puts the next reference on the next line

\def\ref#1#2{\xdef #1{[\number\nref]} % define reference label
\ifnum\nref = 1\global\xdef\therefs{\item{[\number\nref]} #2\ } % first ref
\else% not the first ref
\global\xdef\oldrefs{\therefs}% old reference list
\global\xdef\therefs{\oldrefs\vskip.1in\item{[\number\nref]} #2\ }%
\fi%
\global\advance\nref by 1% advance label
}

\def\listrefs{\vfill\eject\section{References}\therefs}

% ------------------------------------------------------------------
% Footnotes
% ------------------------------------------------------------------

\newcount\nfoot % create counter for footnotes
\global\nfoot = 1 % initialize it

\def\foot#1#2{\xdef #1{(\number\nfoot)} % define footnote label
\hskip -0.2cm ${}^{\number\nfoot}$ 
\footnote{}{\vbox{\baselineskip=10pt
\eightrm \hskip -1cm ${}^{\number\nfoot}$ #2}}
\global\advance\nfoot by 1% advance label
}

% ------------------------------------------------------------------
% Figures
% ------------------------------------------------------------------

\newcount\nfig % create counter for figures
\global\nfig = 1% initialize it
\def\thefigs{} % in case there are no figures but \figurecaptions appears 

\def\figure#1#2{\xdef #1{(\number\nfig)}% define figure label
\ifnum\nfig = 1\global\xdef\thefigs{\item{(\number\nfig)} #2\ }% first figure
\else% not the first figure
\global\xdef\oldfigs{\thefigs}% old figure caption list
\global\xdef\thefigs{\oldfigs\vskip.1in\item{(\number\nfig)} #2\ }%
\fi%
\global\advance\nfig by 1 } % advance label

% this is the old figure definition which is kept to
% keep the macros compatible with older papers
\def\fig#1{\xdef #1{(\number\nfig)}% define figure label
\global\advance\nfig by 1 } % advance label

% ------------------------------------------------------------------
% Tables
% ------------------------------------------------------------------

\newcount\ntab% create counter for tables
\global\ntab = 1% initialize it

\def\table#1{\xdef #1{\number\ntab}% define table label
\global\advance\ntab by 1 } % advance label

% ------------------------------------------------------------------
% Equations
% ------------------------------------------------------------------

\newcount\cflag% create custom flag
\newcount\nequation% create equation counter
\global\nequation = 1% initialize it
\def\eqlabel{(1)}% initialize equation label

% Increment equation counter
\def\nexteqno{\ifnum\cflag = 0% if no custom numbering
\global\advance\nequation by 1% advance number
\fi% end of conditional
\global\cflag = 0% reset custom flag
\xdef\eqlabel{(\number\nequation)}}% define equation label

% Decrement equation counter
\def\lasteqno{\global\advance\nequation by -1% decrease number
\xdef\eqlabel{(\number\nequation)}}% define equation label

% Label equation
\def\label#1{\xdef #1{(\number\nequation)}% define equation name
\ifnum\dflag = 1% if in draft mode
{\escapechar = -1% locally remove "\"
\xdef\draftname{\littlefont\string#1}}% define draft name (small font)
%\xdef\draftname{\tt\string#1}}% define draft name (typewriter font)
\fi}

% Custom label equation
\def\clabel#1#2{\xdef\eqlabel{(\number\nequation #2)}% define custom label
\global\cflag = 1% set custom flag
\xdef #1{\eqlabel}% label equation
\ifnum\dflag = 1% if in draft mode
{\escapechar = -1% locally remove "\"
\xdef\draftname{\string#1}}% define draft name
\fi}

% Completely custom label equation
\def\cclabel#1#2{\xdef\eqlabel{#2)}% define custom label
\global\cflag = 1% set custom flag
\xdef #1{\eqlabel}% label equation
\ifnum\dflag = 1% if in draft mode
{\escapechar = -1% locally remove "\"
\xdef\draftname{\string#1}}% define draft name
\fi}

% Display equation stuff ---------------------------------------------

% End of equation
\def\eeq{}

% Begin displayed unnumbered equation
\def\eqnn #1\eeq{$$ #1 $$}

% Begin displayed numbered equation
\def\eq #1\eeq{
\ifnum\dflag = 0% if not in draft mode
{\xdef\draftname{\ }}% default = no draft name
\fi % end conditional
$$ #1% print equation
\eqno{\eqlabel \rlap{\ \draftname}} $$% print equation number
\nexteqno}% increment equation number

% Print equation number

% Equation array stuff ----------------------------------------------

% End line with equation number
\def\eol{& \eqlabel \rlap{\ \draftname} \crcr% print equation number
\nexteqno% increment equation number
\xdef\draftname{\ }}% clear draft name

% Last eol
\def\eeol{& \eqlabel \rlap{\ \draftname}% print equation number
\nexteqno% increment equation number
\xdef\draftname{\ }}% clear draft name

% End line without equation number
% clear draft name

% Last eol without equation number
% clear draft name

% begin equation array
\def\eqa #1\eeq{
\ifnum\dflag = 0% if not in draft mode
{\xdef\draftname{\ }}% default = no draft name
\fi % end conditional
$$ \eqalignno{ #1 } $$% print equation
\global\cflag = 0}% reset custom flag

% ------------------------------------------------------------------
% Useful abbreviations
% ------------------------------------------------------------------

\def\ie{{\it i.e.\/}}

% ------------------------------------------------------------------
% Journal abbreviations
% ------------------------------------------------------------------

\def\prb#1#2#3{{\it Phys.\ Rev.} {\bf B#1} (19#2) #3}

\def\rmp#1#2#3{{\it Rev.\ Mod.\ Phys.} {\bf #1} (19#2) #3}

% Math parameters -------------------------------------------------

\global\nulldelimiterspace = 0pt

% Math relations ----------------------------------------------------

% Math operations --------------------------------------------------

\def\frac#1#2{{{#1} \over {#2}}\,}  % fraction
\def\hf{{1\over 2}}
\def\nth#1{{1\over #1}}
  % small fraction

  % derivative
  % partial derivative
\def\grad{\nabla}

\def\Asl{\hbox{/\kern-.7500em\it A}} % A slash
\def\Dsl{\hbox{/\kern-.6700em\it D}} % D slash
\def\dsl{\hbox{/\kern-.5300em$\partial$}}
\def\pxpsl{\hbox{/\kern-.5600em$p$}}
\def\sslsh{\hbox{/\kern-.5300em$s$}}
\def\epssl{\hbox{/\kern-.5100em$\epsilon$}}
\def\delsl{\hbox{/\kern-.6300em$\nabla$}}
\def\lxpsl{\hbox{/\kern-.4300em$l$}}
\def\elxpsl{\hbox{/\kern-.4500em$\ell$}}
\def\kxpsl{\hbox{/\kern-.5100em$k$}}
\def\qxpsl{\hbox{/\kern-.5000em$q$}}
 % transpose
\def\sla#1{\raise.15ex\hbox{$/$}\kern-.57em #1}% Feynman slash

% Math accents -----------------------------------------------------

%\def\supsub#1#2{\mathstrut^{\vphantom{\dagger}#1}_{\vphantom{A}#2}}
%\def\sub#1{\mathstrut^{\vphantom{\dagger}}_{\vphantom{A}#1}}
%\def\sup#1{\mathstrut_{\vphantom{A}}^{\vphantom{\dagger}#1}}
%\def\rsub#1{\mathstrut^{\vphantom{\dagger}}_{\vphantom{A}\rm #1}}
%\def\rsup#1{\mathstrut_{\vphantom{A}}^{\vphantom{\dagger}\rm #1}}

\def\roughly#1{\mathrel{\raise.3ex\hbox{$#1$\kern-.75em\lower1ex\hbox{$\sim$}}}}
\def\lsim{\roughly<}

% Alphabets --------------------------------------------------------

% Lower Case Bold Face

% Upper Case Bold Face

% Upper Case Script

% Upper Case ScriptScriptStyle

\def\ssf{{\sss F}}

\def\ssn{{\sss N}}

\def\sst{{\sss T}}

\def\ssx{{\sss X}}

% MohammedÕs general bolf-face macro

\def\pmb#1{\setbox0=\hbox{#1}%
\kern-.025em\copy0\kern-\wd0
\kern.05em\copy0\kern-\wd0
\kern-.025em\raise.0433em\box0}   

% JuanÕs bold-face greek letters
% the template is  \bolded{\XXXX}

\font\jlgtenbrm=cmbx10
\font\jlgtenbit=cmmib10
\font\jlgtenbsy=cmbsy10
\font\jlgsevenbrm=cmbx10 at 7pt
\font\jlgsevenbsy=cmbsy10 at 7pt
\font\jlgsevenbit=cmmib10 at 7pt
\font\jlgfivebrm=cmbx10 at 5pt
\font\jlgfivebsy=cmbsy10 at 5pt
\font\jlgfivebit=cmmib10 at 5pt
\newfam\jlgbrm

\textfont\jlgbrm=\jlgtenbrm
\scriptfont\jlgbrm=\jlgsevenbrm
\scriptscriptfont\jlgbrm=\jlgfivebrm
\newfam\jlgbit

\textfont\jlgbit=\jlgtenbit
\scriptfont\jlgbit=\jlgsevenbit
\scriptscriptfont\jlgbit=\jlgfivebit
\newfam\jlgbsy

\textfont\jlgbsy=\jlgtenbsy
\scriptfont\jlgbsy=\jlgsevenbsy
\scriptscriptfont\jlgbsy=\jlgfivebsy
\newcount\jlgcode
\newcount\jlgfam
\newcount\jlgchar
\newcount\jlgtmp
\def\bolded#1{
% mathchar = (class*16+fam)*256+char (Knuth p. 155)
% class = mathchar/4096
% fam = (mathchar-class*4096)/256
% char = mathchar-(class*16+fam)*256
        \jlgcode\the#1 \divide\jlgcode by 4096
        \jlgtmp\the\jlgcode \multiply\jlgtmp by 4096
        \jlgfam\the#1 \advance\jlgfam by -\the\jlgtmp
        \divide\jlgfam by 256
        \jlgtmp\the\jlgcode \multiply\jlgtmp by 16
	\advance\jlgtmp by \the\jlgfam
	\multiply\jlgtmp by 256
        \jlgchar\the#1 \advance\jlgchar by -\the\jlgtmp
        \advance\jlgfam by \the\jlgbrm
        \jlgtmp\the\jlgcode
        \multiply\jlgtmp by 16
        \advance\jlgtmp by \the\jlgfam
        \multiply\jlgtmp by 256
        \advance\jlgtmp by \the\jlgchar
        \mathchar\the\jlgtmp
}

% Math functions ---------------------------------------------------

\def\det{\mathop{\rm det}}

% Math constructs --------------------------------------------------

% bras 'n' kets

% integral measures

% Abbreviations ----------------------------------------------------

\def\hc{{\rm h.c.}}

% units

\def\eV{{\rm \ eV}}